\documentstyle[pra,aps,twocolumn]{revtex}

\begin{document}
\draft
\title{Output Coupling For an Atom Laser by State Change}
\author{G.M. Moy \cite{email:Moy} and C.M. Savage}
\address{Department of Physics and Theoretical Physics,
The
Australian National University, \\
Australian Capital Territory 0200,
Australia.}
\date{\today}
\maketitle

\begin{abstract}
We calculate the spectrum of a beam of atoms output from a single mode 
atomic cavity.  The output coupling uses an internal state change to 
an untrapped state.  We present an analytical solution for the output 
energy spectrum from a broadband coupler of this type.  An example of 
such an output coupler, which we discuss in detail uses a Raman 
transition to produce a non-trapped state.
\end{abstract}


\narrowtext

As a result of recent experiments in which a Bose Einstein Condensate 
(BEC) has been produced in the 
lab\cite{Anderson95,Bradley95,Davis95,Mewes96} there has been 
considerable interest in coupling the atoms in a BEC out of a trap.  
This could produce a continuous, coherent, directional beam of atoms - 
an atom laser beam 
\cite{Holland95,Wiseman95a,Wiseman95b,Guzman96,Spreeuw95,Olshanii95,Moy97}. 
 While initial experiments have succeeded in coupling atoms out of a 
BEC by changing the internal state of the atoms to a non-trapped 
state\cite{MITExpt1,MITExpt2}, there is still much to be understood 
about the output beam.  In this paper we present an analytical 
solution for the output energy spectrum of atoms in a single trapped 
mode coupled to free space by a change of internal state.  Our 
analysis is based on the atom field input-output theory presented by 
Hope\cite{Hope97}.  We discuss the dependence of the spectrum on 
output coupling strength, and relate these findings to the MIT atom 
laser experiment\cite{MITExpt1,MITExpt2}.

In a BEC a large number of bosonic atoms are cooled into a single 
energy eigenstate of a trap.  This is an important step towards 
producing a monoenergetic beam of atoms.  Nevertheless we still have 
the problem of how to coherently couple the atoms out of such a trap 
in a way that preserves their monoenergetic nature.
 
There are many ways in which atoms can be coupled out of a trap.  The 
simplest method is to turn off the trap\cite{Guzman96}.  The result of 
rapidly turning off the trap is to reproduce the BEC wavefunction in 
free space.  In particular, the wavefunction momentum width is 
conserved.  As a result, the atoms have the corresponding range of 
energies in free space and the monoenergetic nature of the original 
BEC is lost.  Fortunately, energy conserving output coupling is 
possible.  One example is quantum mechanical tunneling of atoms 
through the trap walls.  This is the atomic analogue to the use of 
partially transparent mirrors on an optical laser.  Such a process has 
been considered in a model of an atom laser proposed by 
Wiseman\cite{Wiseman95a}.  It would be difficult in practice, however, 
to use tunneling to produce sufficient fluxes of atoms due to the 
exponential dependence of the tunneling rate on the trap potential 
barrier.

Another approach to the output coupling problem would be to change the 
internal state of the trapped atoms to an untrapped state.  
Experimentally such a method has been used by implementing 
radio-frequency pulses to induce spin flips on trapped atoms in a 
BEC\cite{MITExpt1,MITExpt2}.  Furthermore the use of Raman transitions 
as a method of output coupling has been suggested\cite{Moy97}.  Raman 
transitions have a number of advantages.  A Raman transition can have 
an extremely narrow linewidth so that lasers can be tuned so as to 
only couple atoms from a particular trap mode, due to energy 
conservation.  Moreover when Raman beams are oriented so that they are 
counter propagating, they provide a momentum kick of size $2 \hbar k$.  
This could be used to provide directionality to the atomic output beam 
if atoms were supported against gravity, for instance in a hollow 
optical fiber\cite{Marksteiner93,Ito95,Renn96}.

We model here an output coupler based on change of state, focusing 
initially on the specific case of a Raman output coupler which uses 
two lasers tuned to a two-photon resonance to couple atoms between an 
initial atomic state, and a final atomic state.  There is a third, 
excited, atomic state which mediates the Raman transition.  We assume 
that each of the lasers is far detuned from single photon resonance.  
In this far detuned limit we can adiabatically eliminate the third 
state to produce an effective two level Hamiltonian.  In this 
Hamiltonian we ignore the energies of higher atomic modes of the trap.  
Initially these other modes are empty as we assume all the atoms are 
condensed in the ground mode.  Ignoring these higher energy modes for 
later time is valid for very narrow linewidth Raman lasers which are 
only on resonance with the ground trap mode.  This ensures that higher 
modes do not become populated by atoms in the output state 
transferring back into the initial state at later times.  In addition 
population of other modes is suppressed by Bose enhancement of 
transitions into the ground mode\cite{Moy97}.  We also ignore the 
effects of atom-atom interactions.  The resulting effective 
Hamiltonian is then of the form

\begin{eqnarray}
{\cal H_{\mbox{eff}}} &=& {\cal H_{\mbox{sys}}} + 
      {\cal H_{\mbox{ext}}}  + {\cal H_{\mbox{int}}}, \label{Eq.Heff}\\
{\cal H_{\mbox{sys}}} &=& \hbar \widetilde{\omega}_{0} ~ a^{\dag} a, 
\label{Eq.Hsys}\\
{\cal H_{\mbox{ext}}} &=& \int \mbox{dk} ~ \hbar \widetilde{\omega}_{k} ~ 
      b_{k}^{\dag} b_{k}, \label{Eq.Hext}\\
{\cal H_{\mbox{int}}} &=& -i \hbar \int \mbox{dk}~~ (\kappa (k,t) ~b_{k} 
      a^{\dag} - \kappa^{*} (k,t)~ b_{k}^{\dag} a), \label{Eq.Hint}
\end{eqnarray} 
with
\begin{eqnarray}
\widetilde{\omega}_{0} &=& \omega_{1} + \omega_{0} - 
     \frac{ \Omega_{1}^{2}}{\Delta_{1}}, \label{Eq.w0} \\
\widetilde{\omega}_{k} &=& \omega_{2} + \frac{\hbar k^{2}}{2 m} - 
     \frac{\Omega_{2}^{2}}{\Delta_{2} } ,
     \label{Eq.wk} \\
\kappa (k,t) &=& \Gamma^{\frac{1}{2}}  ~ 
      \left( -i e^{-i(\omega_{2L}-\omega_{1L})t} 
     \psi^{*}(k-k_{1L}-k_{2L}) \right) , \label{Eq.kappa}\\
\Gamma^{\frac{1}{2}} &=& \frac{\Omega_{1} \Omega_{2}}{\Delta_{1}}. 
\label{Eq.Gamma}
\end{eqnarray}
Here, the single trap mode is described by the creation operator, 
$a^{\dag}$ and is coupled by the Raman lasers to a continuous 
spectrum of external modes described by creation operators, 
$b_{k}^{\dag}$.  $\hbar \omega_{1}$ ($\hbar \omega_{2}$) is the energy 
of the trap (output) atomic state.  $\hbar \omega_{0}$ is the ground 
state trap energy.  $m$ is the mass of the trapped atoms.  $\hbar 
k_{1L}$ and $\hbar k_{2L}$ are the momenta of the two lasers inducing 
the Raman transition, with frequencies $\omega_{1L}$ and $\omega_{2L}$ 
respectively.  Thus $\hbar (k_{1L} + k_{2L})$ is the total momentum 
kick received by atoms making the Raman transition.  $\Omega_{1}$ 
($\Omega_{2}$) is the Rabi frequency of the transition between the 
trapped (output) state and the excited state which mediates the Raman 
transition.  $\Delta_{1}$ and $\Delta_{2}$ are the detunings of the 
two Raman lasers from the excited state.  We have assumed these are large in 
adiabatically eliminating the upper level.  If the lasers are tuned 
close to the two-photon resonance, $\Delta_{1} \approx 
\Delta_{2}$.  $\psi(k)$ is the momentum space wavefunction of the 
ground mode of the trap.  $\Gamma$ is a coupling strength, given here 
in terms of the Rabi frequencies and single photon detuning.

The form of the Hamiltonian, Eqs.  (\ref{Eq.Hsys} - \ref{Eq.Hint}), is 
valid in the more general case of an arbitrary output coupling through 
state change involving a single mode system coupled to a continuous 
spectrum of external modes.  In this more general case $\hbar 
\widetilde{\omega}_{0}$ gives the energy of the trapped atoms, and 
$\hbar \widetilde{\omega}_{k}$ gives the energy of the free atoms.  
The coupling strength is more generally defined through $\kappa(k,t) = 
\Gamma^{\frac{1}{2}} \kappa'(k,t)$ where $\kappa'(k,t)$ describes only 
the shape of the coupling and is normalised to unity.  The form of 
$\kappa'(k,t)$ for a general interaction describing a change of state 
is $\kappa(k,t) = \psi^{*}(k-k_{0})$\cite{Hope97}.  Here, $\psi(k)$ is 
the ground state momentum space wavefunction of the single mode system 
and $k_{0}$ describes a possible fixed momentum kick applied to the 
atoms in the state change process.  In the following we discuss the 
Raman coupling case, given by Eqs.  (\ref{Eq.w0} - \ref{Eq.Gamma}) for 
definiteness.  The results, however, are valid for a general output 
coupler in the regime where the coupling strength, $\Gamma$, and the 
energies $\hbar \tilde{\omega}_{0}$ and $\hbar \tilde{\omega}_{k}$ are 
suitably defined.

We are interested in the output energy spectrum, $\langle b_{k}^{\dag} 
b_{k} \rangle$ which is the mean population density of the continuum 
of free space momentum eigenstate modes, labelled by the momentum 
$\hbar k$.  We obtain this by solving the Heisenberg equations of 
motion for the operators, $b_{k}(t)$.  In general, such a solution is 
difficult to obtain, however recently Hope \cite{Hope97} has presented 
a solution in terms of inverse Laplace transforms.  Using these 
solutions, the output spectrum, in the case where initially the 
external modes are empty is given by

\begin{eqnarray}
\langle b_{k}^{\dag}(t) b_{k}(t) \rangle &=& |\kappa(k,t)|^{2} 
       ~~\langle a^{\dag}(0) a(0) \rangle ~~|M_{k}(t)|^{2}, 
       \label{Eq.Spectrum}
\end{eqnarray}
where
\begin{eqnarray}
 M_{k}(t) &=& {\cal L}^{-1} \left\{ \frac{1}{(s + {\cal L}(f')(s)) ~~
      (s + i \delta_{k})}     \right\} (t), \label{Eq.Mkt1}\\
f'(t) &=& \int \mbox{dk} ~ |\kappa(k,t)|^{2} ~ e^{-i \delta_{k} t} ,
\label{Eq.f't}\\
\delta_{k} &=& \widetilde{\omega}_{k} - \widetilde{\omega}_{0} - \omega_{1L} + 
\omega_{2L}  = \frac{\hbar k^{2}}{2 m} - \omega_{0}. 
\label{Eq.deltak}
\end{eqnarray}
The final equality holds for the case when the lasers are tuned to the 
two photon resonance in free space, which we assume here.  ${\cal L}$ 
and ${\cal L}^{-1}$ are the Laplace transform and inverse Laplace 
transform respectively.

We present an analytic solution for the spectrum in the limit of 
broadband coupling.  For simplicity, we consider the case where the 
total momentum kick from the Raman lasers is very small.  That is we 
assume $k_{1L} \approx -k_{2L}$.  This is analogous to the MIT output 
coupling experiments in which the atoms receive a negligible momentum 
kick in changing state\cite{MITExpt1,MITExpt2}.  We also assume that 
the coupling function $\kappa(k,t)$ is broad.  The shape of 
$\kappa(k,t)$ is given by the ground state momentum wavefunction of 
the trap, $\psi(k)$.  We consider here a harmonic trap, with a 
gaussian ground state of standard deviation $\sigma_{k}$ in wavenumber 
space.  We can calculate an exact value for ${\cal L}(f')(s)$ from the 
definition given in Eq.  (\ref{Eq.f't}), however we must simplify 
${\cal L}(f')(s)$ in order to evaluate Eq.  (\ref{Eq.Mkt1}).  In the 
regime where $\mbox{Im}(s) << \hbar \sigma_{k}^{2}/m$ we can 
approximate ${\cal L}(f')(s)$ by
\begin{eqnarray}
{\cal L}(f')(s) &\approx& \Gamma c  \frac{\sqrt{i}}
       {\sqrt{s - i \omega_{0}}}, \label{Eq.Laplacef'} \\
c &=& -i \left( \frac{m \pi}{\hbar \sigma_{k}^{2}} \right)^{1/2} .
\label{c}
\end{eqnarray}
Using this approximation to calculate $M_{k}(t)$ is equivalent to 
discarding high ($> \hbar \sigma_{k}^{2}/m$) frequency information in 
the Laplace transform space.  As we increase the width of our coupling 
in momentum space, given by $\sigma_{k}$, our solution for $M_{k}(t)$ 
becomes valid for increasingly high frequencies.  For an infinitely 
broad coupling our expression becomes exact, and is equivalent to the 
form of the general broadband coupling discussed by Hope\cite{Hope97}.  
Using the above expression for ${\cal L}(f')(s)$ we find the inverse 
Laplace transform, $M_{k}(t)$ to be

\begin{eqnarray}
        M_{k}(t) & = & -e^{i \omega_{o} t} \frac{i \sqrt{i} \Gamma c}
          {\left(\omega_{k} \Delta_{k}^{2} - \Gamma^{2} c^{2}\right) 
          \sqrt{\pi t}} \nonumber  \\
         &  &+ e^{-i\Delta_{k}t} \frac{i \omega_{k} \Delta_{k}}
          {\omega_{k} \Delta_{k}^{2} - \Gamma^{2} c^{2}}
        \nonumber  \\
         &  &+  e^{-i\Delta_{k}t} \frac{1}{2} \sqrt{\frac{\pi}{t}} \frac{i \sqrt{i} \Gamma c}
          {\omega_{k} \Delta_{k}^{2} - \Gamma^{2} c^{2}}
             L^{-1/2}_{1/2}(i\omega_{k}t)
        \nonumber  \\
         &  & + \frac{\alpha^{2}\exp \left[(\alpha^{2} + i \omega_{0})t 
         \right]}{(\beta-\alpha)(\gamma-\alpha)
                            (\alpha^{2}+i\omega_{k})} (1+\mbox{Erf}(\alpha 
                            \sqrt{t}))\nonumber \\
         &  & + \frac{\beta^{2}\exp \left[(\beta^{2} + i \omega_{0})t
         \right]}{(\alpha-\beta)(\gamma-\beta)
                            (\beta^{2}+i\omega_{k})} (1+\mbox{Erf}(\beta 
                            \sqrt{t}))\nonumber \\
         &  & + \frac{\gamma^{2}\exp \left[(\gamma^{2} + i \omega_{0})t
         \right]}{(\alpha-\gamma)(\beta-\gamma)
                            (\gamma^{2}+i\omega_{k})} (1+\mbox{Erf}(\gamma \sqrt{t})),
         \label{Eq.Mk(t)}
\end{eqnarray}
where we have defined $\omega_{k} = \hbar k^{2}/(2 m)$ and $\Delta_{k} 
= \omega_{k} - \omega_{0}$.  The function $L_{\frac{1}{2}}^{- 
\frac{1}{2}}(x)$ is a Laguerre polynomial, Erf is the error 
function and $\alpha$, $\beta$ and $\gamma$ are the roots of the 
equation $s^{3} + i \omega_{0} s + \Gamma c \sqrt{i} = 0$.

Fig.  \ref{Fig.Mkt} shows the behaviour of $|M_{k}(t)|^{2}$ as a 
function of $\omega_{k}$ and time after we turn on the output coupling 
interaction.  Initially $|M_{k}(t)|^{2}$ is small, and for short 
enough times, arbitrarily broad in k-space.  Initially 
$|M_{k}(t)|^{2}$ agrees with the perturbative solutions presented by 
Hope \cite{Hope97}.  For longer times, we can see that the spectrum 
reaches a stable shape.  For very large values of the coupling 
strength, the long time limit becomes very broad in k-space.  As a 
result, the shape of the output spectrum, as given by Eq.  
(\ref{Eq.Spectrum}), simply reflects the momentum distribution of the 
cavity wave-function, $\psi(k)$.  As a result there is no narrowing of 
linewidth in momentum space.  The recent MIT 
experiments\cite{MITExpt1,MITExpt2} are an example of an output 
coupling with an extremely large coupling strength.  In these 
experiments a short, $5 \mu \mbox{s}$ RF pulse was used to couple 
atoms out of a BEC, making a pulsed atom laser.

We consider here a continuous coupler, turned on at time $t=0$, and 
examine the resulting long time spectrum in the external modes 
described by $b_{k}^{\dag}$.  We observe in Fig.  \ref{Fig.Mkt} that 
for longer times $|M_{k}(t)|^{2}$ narrows into a sinc function centered 
about the trap ground state frequency, $\omega_{0}$.  Eventually 
$|M_{k}(t)|^{2}$ reaches a stationary state with a lorentzian like profile 
as shown in Fig.  \ref{Fig.Mkt}.  This longtime behaviour is given by
\begin{eqnarray}
\lim_{t \rightarrow \infty} M_{k}(t) &=& \frac{i \sqrt{\omega_{k}} e^{-i 
\Delta_{k} t}}{\sqrt{\omega_{k}} \Delta_{k} - \Gamma c} \nonumber\\
&&~\mbox{          }+
\frac{2 \gamma^{2} e^{i(\omega_{0} + \gamma^{2}) t}}{ (\alpha - 
\gamma) (\beta - \gamma) (\gamma^{2} + i \omega_{k}) }, \label{Eq.longtime}
\end{eqnarray}
where $\gamma$ is the particular solution to the cubic discussed above, given by 
the expression
\begin{eqnarray}
\gamma &=& e^{i \frac{\pi}{4}} \left(\frac{2^{\frac{1}{3}} 
\omega_{0}}{\xi^{\frac{1}{3}}} - \frac{\xi^{\frac{1}{3}}}{3 
2^{\frac{1}{3}}} \right), \\
\xi &=& -27 i \Gamma c + \left( (27 \Gamma c)^{2} + 108 
\omega_{0}^{3} \right)^{\frac{1}{2}}. \nonumber
\end{eqnarray}
The longtime expression for $M_{k}(t)$, Eq.  (\ref{Eq.longtime})
contains two terms.  The first of these terms dominates in the case of 
small $\Gamma$, while the second dominates for very large $\Gamma$.  
As a result, the long time spectrum has two distinct behaviours 
depending on the strength of the coupling.  We consider the case of 
slow coupling (small $\Gamma$) initially.  In this case, the long time 
expression for $M_{k}(t)$ is dominated by the first term in Eq.  
(\ref{Eq.longtime}) above, and the resulting long time spectrum is 
given by
\begin{eqnarray}
\langle b_{k}^{\dag} b_{k} \rangle &=& \Gamma ~|\psi(k)|^{2} ~
\frac{1}{\left( \Delta_{k}^{2} + |\Gamma c|^{2}/\omega_{k} \right)}.
\label{Eq.longtimespectrum}
\end{eqnarray}
A plot of the long time spectrum, Eq.  (\ref{Eq.longtimespectrum}) as 
a function of $\omega_{k}$ is presented in Fig.  \ref{Fig.longtime1} 
for various coupling strengths.  Fig.  \ref{Fig.longtime1} shows that 
for increasing coupling strength the linewidth of the long time 
spectrum increases.  The values for $\Gamma$ chosen correspond 
approximately to values of Raman laser Rabi frequencies, $\Omega_{1} 
\approx 2 \pi \times ~50~ \mbox{kHz}$ and $\Omega_{2} \approx 2 \pi~ 
\times 1.6 ~ \mbox{MHz}$ and detuning, $\Delta_{1} \approx 2 \pi \times 
~2.5 ~\mbox{GHz}$ similar to values presented in \cite{Moy97}.  
However, much smaller or larger coupling strengths can be achieved by 
suitably adjusting the intensities of the lasers and their detunings.  
The figures assume a trap with ground state frequency $\omega_{0} = 2 
\pi~\times 123~ \mbox{s}^{-1}$, typical of magnetic traps for ultra 
cold atoms\cite{Trapw0Ref}.  A ground state gaussian with width 
$\sigma_{k} \approx 10^{6} \mbox{m}^{-1}$ has been assumed, which 
corresponds to a position space wavefunction of size of the order of 
$2 \mu \mbox{m}$.  This value of $\sigma_{k}$ corresponds to a width 
in $\omega_{k}$ space of $\sigma_{\omega_{k}} \approx 10^{4} 
\mbox{s}^{-1}$.

For each of the graphs shown in Fig.  \ref{Fig.longtime1}, the 
lorentzian like spectrum is centred about $\omega_{0}$, the ground 
state frequency of the single mode trap, with the width of the 
spectrum dependent on the strength of the coupling as mentioned above.  
In all cases, however, the linewidth is much less than that which 
would be obtained if the trap was rapidly turned off, that is 
$\sigma_{\omega_{k}} \approx 10^{4} \mbox{s}^{-1}$.  We see from Eq.  
(\ref{Eq.longtimespectrum}) that the distribution isn't exactly 
lorentzian due to the presence of $\omega_{k}$ in the second part of 
the denominator.  However for large $\omega_{0}$ the spectrum is well 
approximated by a lorentzian distribution of width $|\Gamma c|/ 
\sqrt{\omega_{0}}$.

We have already noted that for large coupling rates, the width of the 
longtime limit of $|M_{k}|^{2}$, and hence of the longtime spectrum is 
increased.  When $\Gamma$ is very large, $|\Gamma c|/\sqrt{\omega_{0}} >> 
\sigma_{\omega_{k}}$, the width of $M_{k}(t)$ becomes large compared 
with $\kappa(k,t)$ and the spectrum becomes dominated by the cavity 
momentum spread $\psi(k)$.  As a result, for sufficiently fast 
coupling (large $\Gamma$) the output spectrum changes significantly 
from the lorentzian shape considered above, and instead reflects the 
momentum spread of the cavity.  This is shown in Fig.  
\ref{Fig.longtime2}.  For very large $\Gamma$ the spectrum is centred 
about zero, and falls away exponentially in $\omega_{k}$ space, as 
required for a gaussian distribution in momentum space given by 
$\psi(k)$.

We have shown that the longtime spectrum from an output coupler based 
on state change depends on the strength of the output coupling.  For 
very strong coupling, the output spectrum is given by the cavity 
spectrum, and is very broad in momentum space.  The spectrum is then 
centered about the zero of momentum when there is no net momentum kick 
from the lasers.  As the strength of the coupling is reduced, however, 
the long time linewidth is correspondingly reduced.  For small 
coupling strengths the final linewidth is effectively lorentzian, 
centred about the energy of the cavity with a linewidth proportional 
to the coupling strength $\Gamma$.

The authors would like to thank Joseph Hope for much advice and many 
thoughtful discussions.

\begin{figure}
\caption{Plot of $|M_{k}(t)|^{2}$ as a function of $\omega_{k}$ and 
time for $t=0 \mbox{s}$ to $t=5 \mbox{s}$, and $\omega_{k}$ ranging from 
$762 \mbox{s}^{-1}$ 
to $783 \mbox{s}^{-1}$ about the single mode trap frequency, $\omega_{0} 
\approx 772 \mbox{s}^{-1}$.  $\Gamma = 1.8 \times 10^{3} 
\mbox{s}^{-2}$}
\label{Fig.Mkt}
\end{figure}

\begin{figure}
\caption{Plot of the long time behaviour of $\langle 
b_{k}^{\dag}b_{k} \rangle$ as a function of $\omega_{k}$ for various 
coupling strengths, $\Gamma = 10^{4} \mbox{s}^{-2}$ (dotted line), $\Gamma = 3 
\times 10^{4} \mbox{s}^{-2}$ (solid line) and $\Gamma = 5 \times 10^{4} 
\mbox{s}^{-2}$ 
(dashed line).}
\label{Fig.longtime1}
\end{figure}

\begin{figure}
\caption{Plot of the steady state behaviour of $\langle 
b_{k}^{\dag}b_{k} \rangle$ as a function of $\omega_{k}$ for the large 
coupling limit ($\Gamma \approx 10^{13} \mbox{s}^{-2}$).} \label{Fig.longtime2}
\end{figure}

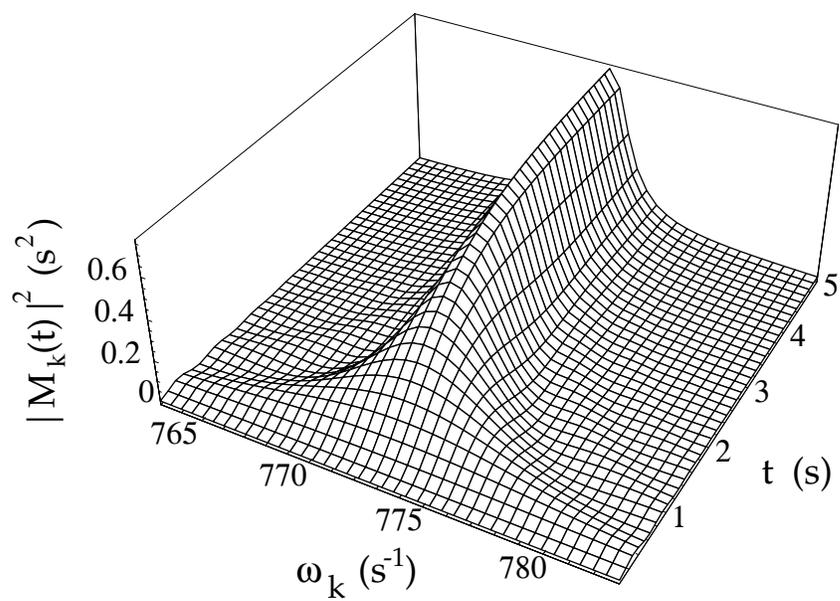
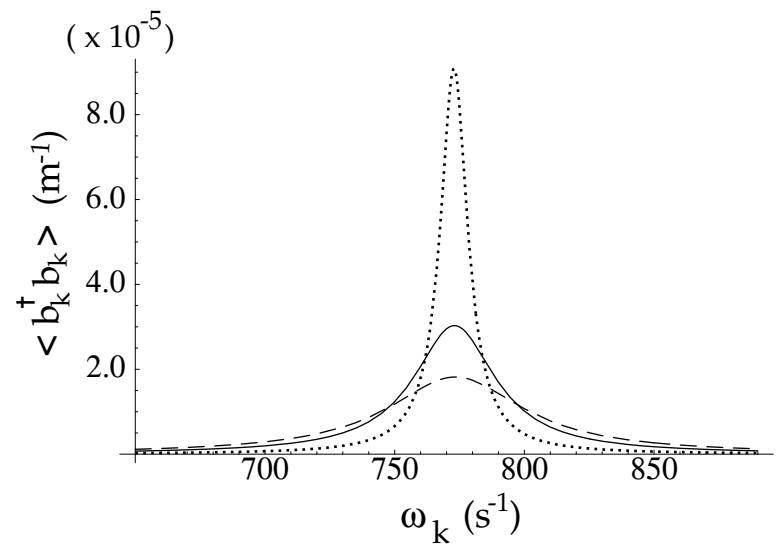
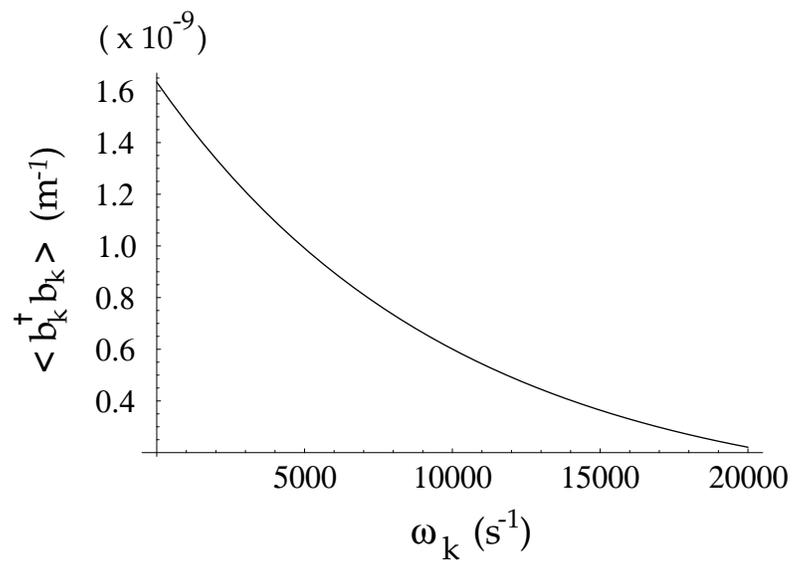
\end{document}